\documentclass[preprint,proceedings]{rmaa}

\usepackage{paralist}

\usepackage{psfrag,color}

\SetYear{2004}
\SetConfTitle{IAU Coll. No. 194: Compact Binaries in the Galaxy and Beyond}

\title{Symbiotic Stars in the Magellanic Clouds} 

\author{
  Joanna Miko{\l}ajewska\altaffilmark{1} 
  }

\altaffiltext{1}{N. Copernicus Astronomical Center, Warsaw, Poland}

\shortauthor{Miko{\l}ajewska}
\shorttitle{Symbiotic Stars in the MCs}

\fulladdresses{
\item N. Copernicus Astronomical Center, Bartycka 18, 00-716 Warsaw, Poland
  (\email{mikolaj@camk.edu.pl}).
}

\listofauthors{J. Miko{\l}ajewska}
\indexauthor{J. Miko{\l}ajewska}

\abstract{Orbital periods and other parameters of symbiotic binary systems 
in the LMC and SMC are presented and discussed. In particular, the symbiotic 
stars in the MCs are compared with those in the Milky Way.}

\addkeyword{Magellanic Clouds}
\addkeyword{Stars: symbiotic}
\addkeyword{Stars: binary systems}
\addkeyword{Stars: parameters}

\begin{document}
\maketitle

Symbiotic stars are long-period interacting binary systems in which 
an evolved giant star (normal giant in S-types, and Mira variable surrounded by thick dust envelope in D-types, respectively) transfers material to its much hotter compact companion.
Such a  composition places them among 
the intrinsically brighest variable stars that can be easily detected in nearby
galaxies, in particular in the Magellanic Clouds (MC).\,At present the number of
confirmed symbiotics is at 6 in the SMC and 8 in the LMC (e.g.\,Belczy{\'n}ski
et al.\,2000).
Here, I present and discuss some physical parameters of MC symbiotic stars derived from available observational data, in particular, those contained in MACHO, 2MASS,
IUE and HST databases.
 
\begin{figure}[!t]
  \includegraphics[width=\columnwidth]{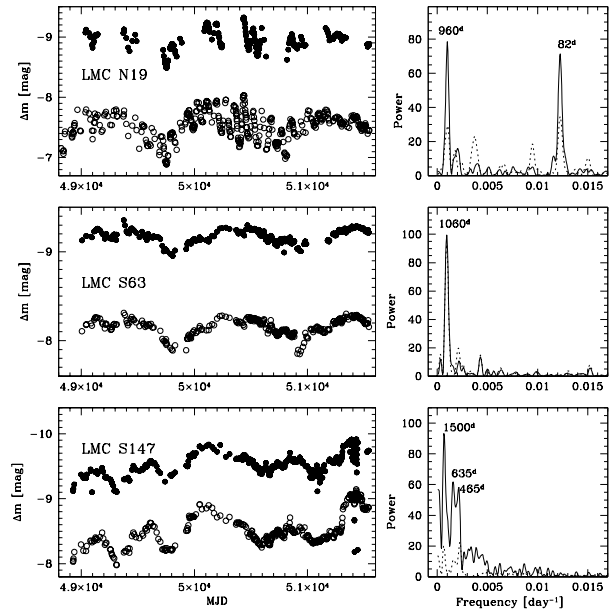}
    \caption{MACHO light curves for LMC N19 (top), LMC S63 (middle), 
and LMC S147 (bottom), and the corresponding power spectra.\,The blue and red channel data are plotted as open and filled circles (light curves), and solid and dashed lines (periodograms), respectively.}
  \label{fig:lc}
\end{figure}

The MACHO/OGLE II light curves of SMC 3 have been recently discussed
by Kahabka (2004) who found evidence for $\sim 1600^{\rm d}$ orbital modulation,
in addition to $110^{\rm d}$ pulsation of the M0 giant component.
The MACHO light curves for LMC systems, S63, N19 and S147 are shown in 
Figure~\ref{fig:lc} together with the corresponding power spectra.\,In all cases there is a strong periodicity of the order of $1000^{\rm d}$, which 
can be due to orbital motion.\,
For LMC N19, strong variability in U light was reported by Morgan (1996) who
also gave dates both when the star was was faint and bright.\,Combining
this information with our light curves, we find that epochs of relative faintness reported by Morgan and the deep minima in the light curves
follow the ephemeris: $\rm MJD\,Min=50\,632+946 \times E$.
Similarly, the IUE fluxes in LMC S63 observed in 1982 (maximum of the $1060^{\rm d}$
periodicity) are higher than those in 1994 (ingres) as expected for orbital
modulation.
For LMC S147, the situation is more complicated, and additional observations
(U photometry and/or spectroscopy) are needed to confirm orbital origin 
of the modulation visible in its light curves.
In N19 there is also strong periodicity at $\sim 80^{\rm d}$ which may be due to radial pulsation of the M4 giant component, similar to that found in SMC\,3 (Kahabka 2004). 

\begin{figure*}[!t]
   \includegraphics[height=5cm]{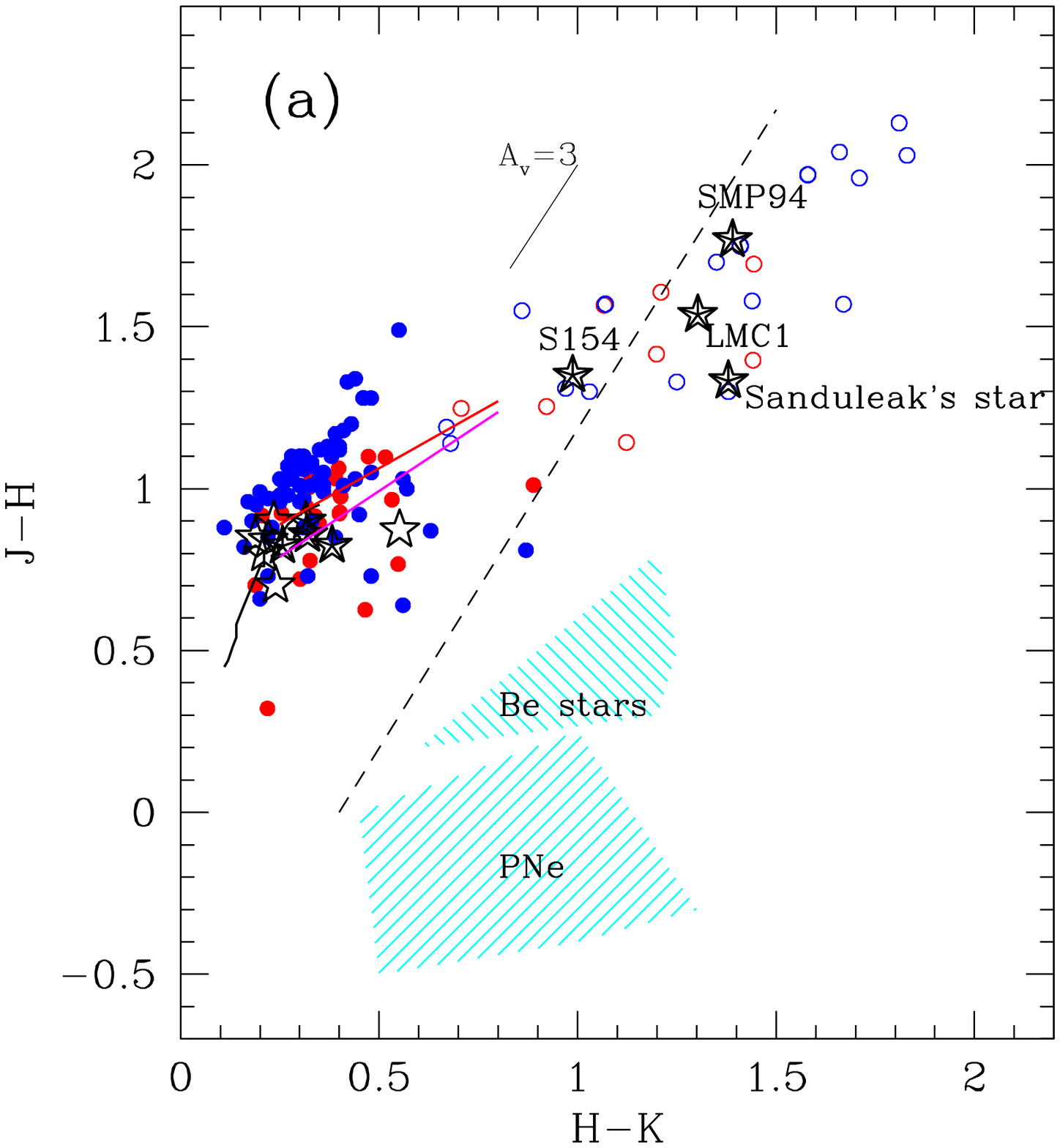}%
  \hspace*{0.5cm}%
  \includegraphics[height=5cm]{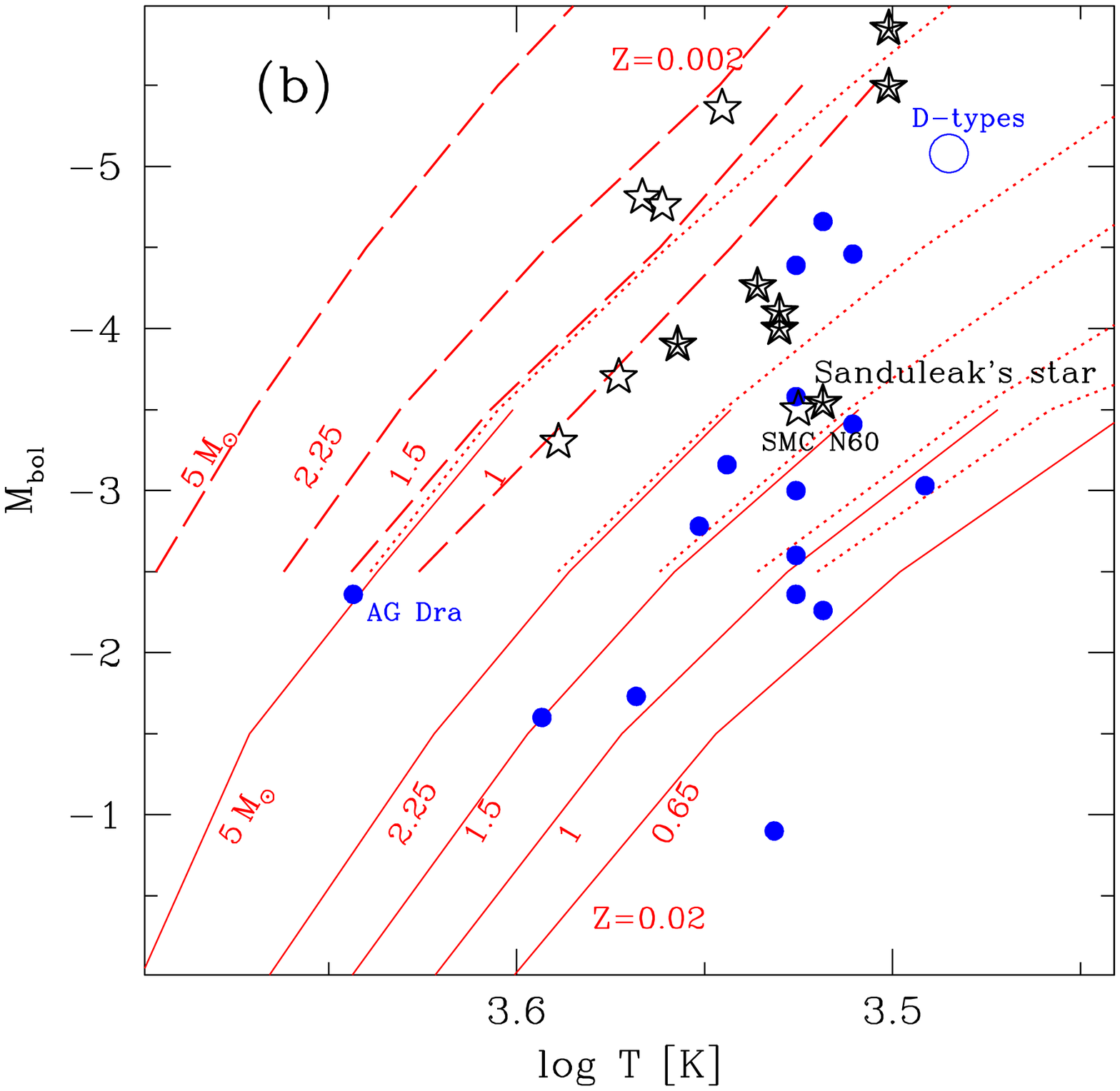}%
\hspace*{0.5cm}%
 \includegraphics[height=5cm]{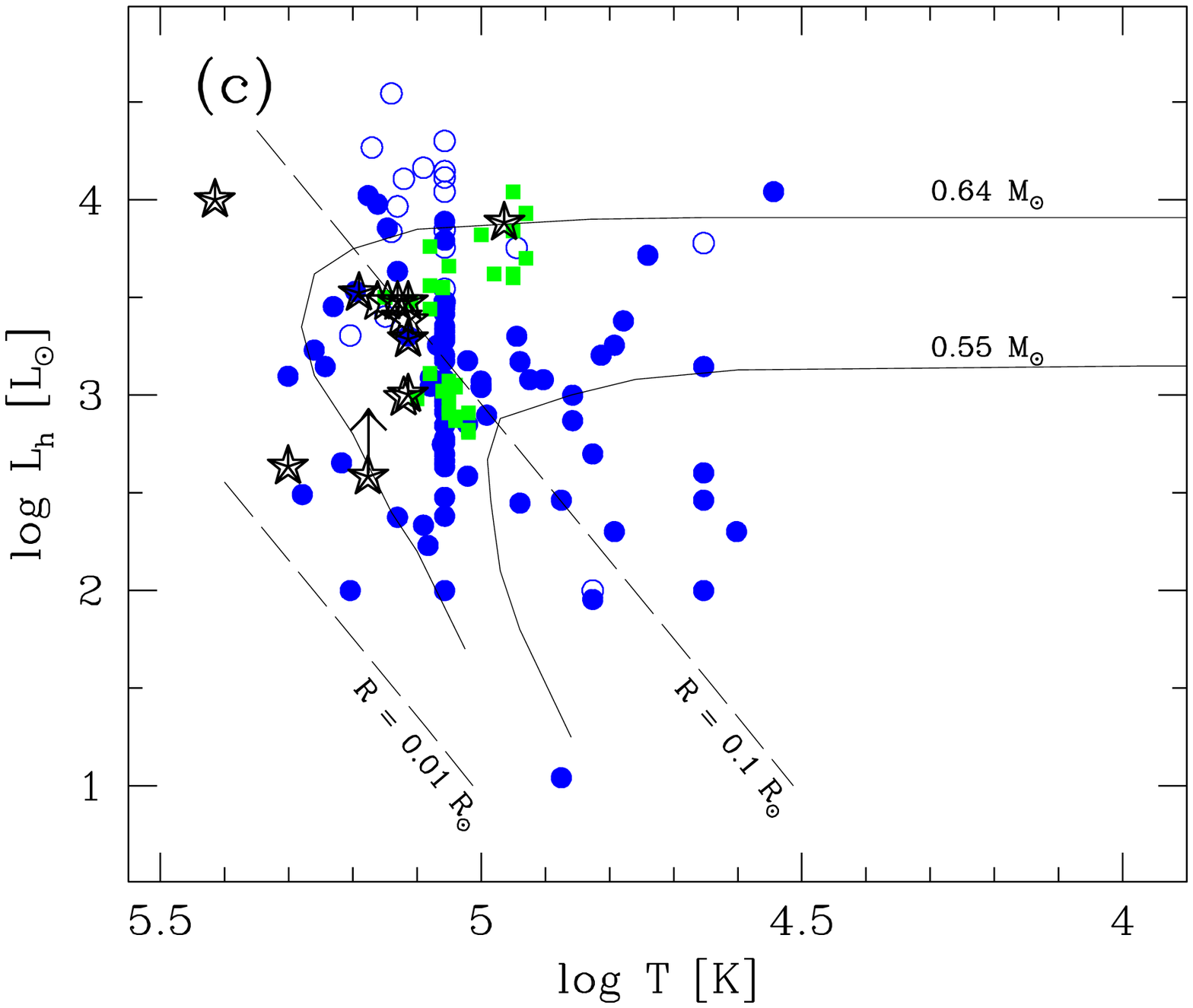}
  \caption{(a) Color-color diagram for MC symbiotic stars from 2MASS 
(open stars correspond 
to the SMC and filled stars to the LMC systems, respectively).\,
For comparison, galactic systems from Munari et al.\,(1992) and 2MASS are also
plotted (closed and open circles represent S- and D-type systems, respectively).
The shaded regions denote planetary nebulae (PNe) and Be stars.\,
Full lines mark the locii of luminosity class III stars and LPVs, and dashed line represent 
black bodies.\,
(b) H-R diagram for the symbiotic giants. Symbols for galactic systems (Miko{\l}ajewska 2004; only system with measured red giant radii, and thus with accurate distances, are plotted) and MC systems are the same as in (a). Evolutionary tracks (Hurley et al.\,2000) for low mass AGB (dashed) and RGB (solid) stars for different Z are also plotted.\,Note the position of the very low Z  galactic system AG Dra in the region occupied by the SMC systems.\, 
(c)\,H-R diagram for the hot components. Galactic (Miko{\l}ajewska 2004) 
and MC systems (M{\"u}rset et al.\,1996; this study) are plotted as circles and stars, whereas whereas filled and open symbols represent the S- and D-systems, 
respectively. Squares correspond to data for AG Dra.
Evolutionary tracks for the cooling white dwarfs (Sch{\"o}nberner 1989), and constant radii are plotted as solid and dashed lines, respectively.}
  \label{fig:fig2}
\end{figure*}

2MASS colors of MC systems and the H-R diagrams for the red giants  
are shown in 
Figure~\ref{fig:fig2}a,b.\,
MC systems contain low mass, $\leq 3\, M_\sun$, giants as do their galactic cousins, however only AGB giants are found in MC systems.\,
Four D-type systems in the LMC of 8 total, wheras no a D-type system in the SMC are found.\,The later result  is surprising because non dusty, S-type galactic systems rarely contain AGB giants.\,The lack of dusty systems among the SMC symbiotics
may reflect the very low $Z$ in the SMC, too low to form enough dust in the giant 
envelope.

The hot components of the galactic and MC systems overlap in the H-R diagram
(Figure~\ref{fig:fig2}c)
but MC systems are among the hottest and the brightest.
This together with the fact that all MC systems contain AGB giants is probably
due to that only the brightest symbiotic systems
have been detected in the MCs.\,It is, however notable that the hot component of AG Dra, the galactic symbiotic with the lowest measured $Z
\sim 0.002$ is also among the hottest systems.\,Further studies are necessary
to study the effect of metallicity on the symbiotic appearance.

Summarising, the first orbital periods for MC symbiotic systems have been found,
all in the range 900--1600 days which is consistent with the longer period tail of galactic S-type systems (Miko{\l}ajewska 2003).\,MC symbiotics are low mass
systems as their galactic cousins.\,Although the MC sample is biased towards the hottest and the brightest systems there 
\adjustfinalcols 
is some evidence for nonegligible metallicity effect on the symbiotic phenomenon.
 
\acknowledgements
I gratefully acknowledge the financial support from the LOC, IAU, and KBN Research Grant No.\,5P03D\,019\,20.\,This study made use of the public domain 
data of the MACHO Project, which was
performed under joint auspices of the DOE, NNSA by the U.\,of Cal., LANL, and the Mt.\,Stromlo and Siding  Spring Obs., and the 2MASS, which is a
joint project of the U.\,of Mass.\,and the IPAC/Caltech, funded by the NASA and the NSF.

\end{document}